\documentclass[aps,prl,twocolumn,showpacs]{revtex4}

\usepackage{graphicx}

\begin{document}
                                                                                                                                                             
\title{A Mechanism for Photoinduced Effects In Tetracyanoethylene-Based 
Organic Magnets}

\author{Serkan Erdin and Michel van Veenendaal\\
Department of Physics, Northern Illinois University, DeKalb, IL, 60115\\
\& Advanced Photon Source, Argonne National Laboratory,\\
 9700 South Cass Avenue,  Argonne, IL, 60439}
                                                                                                                                                             

\begin{abstract}
The photoinduced magnetism in manganese-tetracyanoethylene (Mn-TCNE) 
molecule-based  magnets is ascribed to charge-transfer excitations
 from manganese 
to TCNE. Charge-transfer energies 
are calculated using Density Functional Theory; 
photoinduced magnetization is described using a model Hamiltonian
based on a double-exchange mechanism. 
Photoexciting electrons from the manganese core spin into
the lowest unoccupied orbital of TCNE with  photon energies around 3 eV 
increases the magnetization through a reduction
of the canting angle of the manganese core spins
for an average electron density on TCNE
less than one. When photoexciting
with a smaller energy, divalent TCNE molecules are formed. The delocalization
of the excited electron causes a local spin flip of a manganese core spin. 
\end{abstract}

\maketitle 


In recent years, optical control of magnetic properties  has drawn a 
great 
deal of attention \cite{zutic}. Photoinduced 
changes in magnetic order   were extensively studied in a variety of 
systems, including 
 spin-crossover complexes \cite{exp1}, magnetic heterostructures \cite{exp2},
and 
manganite films \cite{exp3}. Photoinduced effects 
were recently reported  in molecule-based compounds, such as Co-Fe Prussian 
blue  anologs \cite{cofe1,cofe2} and manganese-tetracyanoethylene,  
Mn(TCNE)$_x \cdot y $(CH$_2$Cl$_2$) 
(x $\approx$ 2, y $\sim$ 0.8 ) 
\cite{mntcne}.
In these compounds, magnetization induced by illumination with 
visible light can be partially eliminated by using light 
with a  longer wavelength.   
In the Prussian blue anologs, the increase in  magnetization
is attributed to a low-spin to high-spin transition that is
triggered by photoinduced charge transfer \cite{theorycofe}.
In the case of   Mn(TCNE)$_x \cdot y $(CH$_2$Cl$_2$),
which is the focus of this Letter, optical spectroscopy \cite{mntcne} 
suggests that the photoinduced magnetization can be related to a 
  $\pi \rightarrow \pi^*$ optical transition in TCNE,
whereas a charge transfer between metal and ligand
causes a decrease in magnetization. Despite the enormous 
technological potential of these organic-based molecular magnets
with reversible photoinduced magnetization,
little theoretical work has been done to describe
the exact photoinduced states and the mechanism triggering photoinduced
changes.

Mn(TCNE)$_x \cdot y $(CH$_2$Cl$_2$) is an electron transfer salt in which 
both Mn ion and the organic molecule TCNE carry spins;
CH$_2$Cl$_2$ is a solvent. According to 
M\"ossbauer spectroscopy
studies, the transition-metal ion is divalent with 
manganese  in a high-spin state \cite{mossbauer}. 
The cyanocarbon acceptor TCNE$^-$  has an unpaired electron 
 in the $\pi^*$ molecular orbital. 
The Mn ion  and TCNE molecule are coupled antiferromagnetically with 
a critical temperature of 75 K \cite{reentrant}. At 2.5 K, the magnet
exhibits a transition to a spin glass-like state \cite{reentrant}.
Although the exact structure of the compound is unknown to date, 
experimental evidence suggests that
Mn-TCNE forms a three-dimensional  network in which each Mn ion is 
surrounded by up to six TCNE molecules \cite{struct}.  
The lack of sufficient information on the structure limits the use of 
{\it ab initio} quantum chemistry calculations on 
this magnet. 

Experimentally, it has been observed that the magnetization of the compound 
increases upon argon laser excitation 
in the region $\sim 2.54-3.00$ eV and reaches 
saturation in $6$ h, while the excitation in the region $\sim
1.8-2.5$ eV results in  partial reduction of the
photoinduced magnetization. 
The photoinduced effects persist 
for several days at low temperature. Heating the molecular magnet
up to $200$ K after illumination of light does not fully erase the 
photoexcited state although
photoinduced effects are not observed above $75$ K. 
This implies that magnetic exchange alone does not 
explain the metastable state. 
In this Letter, we show that the excitation energies for photoinduced
magnetization and demagnetization correspond
to electronic transitions between Mn and TCNE with different valencies.
Using Monte-Carlo simulations for a model Hamiltonian, 
we demonstrate that the  magnetic interactions between the manganese spins 
depend dramatically on the valency of the TCNE molecule, explaining 
the changes in magnetization resulting from the photoinduced charge transfers.

\begin{figure}[b]
\begin{center}
\includegraphics[angle=0,totalheight=2.0in]{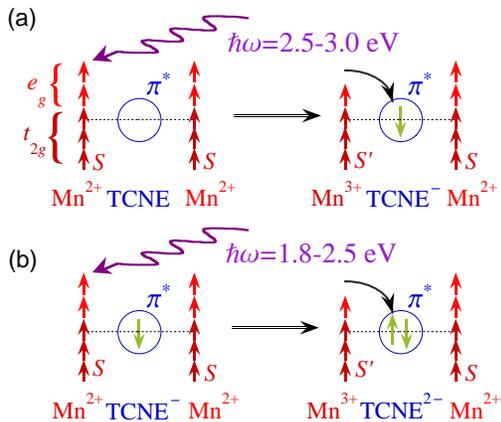}
\caption{\label{mechanism}
(a) The charge-transfer process made for a photon energy of 
$\hbar\omega=2.5$-3.0 eV. An $e_g$ electron from the Mn $S=5/2$, which consists
of 3 $t_{2g}$ electrons and 2 $e_g$ electrons, is transferred to the
$\pi^*$ orbital, the lowest unoccupied state of the TCNE molecule,
leaving a core spin with $S'=2$.
(b) For lower incident photon energies, electrons are transferred 
to a TCNE$^-$ molecule giving rise to a doubly occupied $\pi^*$ orbital. }
\end{center}
\end{figure}

First, we need to understand the nature of the transitions made for different 
excitation energies.
To this end, we performed unrestricted density functional
calculations using GAUSSIAN 03 software \cite{gaussian} with 
local spin density approximation \cite{lsda}. The polarizable-conductor 
calculation model was used \cite{pcm}, 
to take into account the  effects of the solvent. For the TCNE molecule, 
Dunning's correlation-consistent double-$\zeta$ basis set was taken, and an 
effective core potential (LANL2) was used for the core electrons of the Mn 
atom. Calculations for a Mn and TCNE cluster show
 that the excitation energies used experimentally 
correspond to excitations from the Mn $e_g$ states to the TCNE molecule.
In the ground state, 
manganese has
a $t_{2g\uparrow}^3e_{g\uparrow}^2$ configuration with a spin
$S=5/2$ and is octahedrally coordinated by TCNE molecules. 
 After photoexcitation, the Mn ion will be in a trivalent
$t_{2g\uparrow}^3e_{g\uparrow}$ state with $S'=2$, see Fig. \ref{mechanism}(a). 
 The energy for the excitation   Mn$^{2+}$ TCNE $\rightarrow$
Mn$^{3+}$ TCNE$^{-}$ is  $3.18$ eV. The presence of neutral organic molecules in
a strongly disordered compound such as Mn-TCNE is likely
 since some TCNEs can be disoriented  and uncoupled 
to the Mn ions due to the solvent. 
The calculation slightly overestimates the transition energies
due to a lack of band effects because of the finite size of the cluster
and simplifications in the description of the 
polarizability of the solvent. The photoexcited state is expected to be mostly stable
since the Mn$^{3+}$ ion is  Jahn-Teller active 
and the charge transfer induces a distortion of the TCNEs surrounding
the Mn ion. Experimentally, it has been seen that the photoinduced magnetization
is accompanied by lattice distortions \cite{pimpolaron}.
Because the polaronic state does not
depend on the magnetic order, it also explains
why the photoinduced excited state is not fully erased when the
temperature is raised above the critical temperature. 
Relaxation is further complicated by the fact that the $e_g$ states 
couple only weakly to the TCNE valence states, which primarily  consist  
of $\pi$-bonding orbitals.
Photoinduced demagnetization is ascribed to the transition
Mn$^{2+}$  TCNE$^{-}\rightarrow$ Mn$^{3+}$
TCNE$^{2-}$, see Fig. \ref{mechanism}(b),
 which occurs at 2.87 eV. The energy for this transition is lower
due to the electrostatic interaction between the Mn and the TCNE.
Therefore, when tuning the magnetization by illumination with 
visible light, effectively one changes the 
number of valence electrons  by transferring  electrons from the 
 Mn core spin to the lowest unoccupied orbital on TCNE. 
By varying the  wavelength of the light, one can create
different oxidation states of the TCNE that, due to their 
different coupling to the manganese core spins, allows a decrease and
increase in magnetization .

In order to describe how the photoinduced 
 charge transfers between manganese and TCNE can have 
a strongly different impact on the magnetization, we use a model that 
captures the magnetic couplings between the manganese core spins
and perform Monte-Carlo calculations to determine the magnetic
properties.
Of the TCNE, we retain the $\pi^*$ orbital, which is the 
lowest unoccupied molecular orbital. We treat the manganese core spins 
classically. This approach is similar 
to the double-exchange mechanism \cite{de1,de2,de3} often 
used in the colossal magnetoresistive manganites. The Hamiltonian
is given by
\begin{eqnarray}
H_0 &=& - t \sum_{ij\sigma} 
\sin  \frac{\theta_{j\sigma}-\theta_i}{2}  ( 
d_i^\dagger c_{j\sigma} +c_{j\sigma}^\dagger d_i  ) 
 \nonumber \\  &+&
\sum_i \varepsilon_d d_i^\dagger d_i \label{ham}+
\sum_j \varepsilon_{\pi^*}  c_{j \sigma}^\dagger  c_{j \sigma} 
+ J_{AF} \sum_{ii'}{\bf S}_i \cdot {\bf S}_{i'},
\nonumber
\end{eqnarray}
where $c_{i \sigma}^\dagger$ creates an electron
 with spin $\sigma$ on a $\pi^*$ orbital 
of TCNE and $d_j^\dagger$ creates an electron on the 
 effective $d$ orbital of the Mn  site. 
The latter operator  is not assigned a spin, since
 the spin of the conduction electrons
is always antiparallel to that of the manganese core spin. 
Manganese and TCNE  sites are labelled 
by $i$ and $j$, respectively. 
The hopping amplitude between manganese and the $\pi^*$ orbital is
 $t$. $\theta_{i\sigma}$ is the angle of an electron on $i$'th ligand,
which is $0$ or $\pi$ for 
$\sigma=\uparrow$ and $\downarrow$, respectively; 
$\theta_j$ is the angle of classical manganese core spin 
at $j$'th site. 
$\varepsilon_d$ and $\varepsilon_{\pi^*}$ 
 are the energies for $d$ and $\pi^*$ orbitals, respectively.
 The last term in $H_0$ is a superexchange interaction between
two core spins mediated through the highest occupied molecular orbital 
of TCNE. 
We also include the Coulomb interaction on the TCNE molecule,
\begin{eqnarray}
H_U=\sum_j Un_{j\uparrow}n_{j\downarrow}, 
\end{eqnarray}
where $U$ is
the strength of the Coulomb interaction between two electrons in a
$\pi^*$ orbital.
In our calculations, we first construct
the Hamiltonian matrix for $H_0$ 
for a given core spin configuration in a lattice. 
Since the structure is as yet unknown,
we carry out our calculations on a two-dimensional lattice. In doing so, we both
preserve the stoichiometry and use the two-dimensional symmetry of TCNE molecule.
A unit cell contains one
metal site at site $( 0,0 )$ and two ligands at sites
 $( a/2,0 )$ and $( 0,a/2 )$, where $a$ is the distance between two 
manganese atoms.
For our calculations, we choose a lattice of $4\times 4$ unit cells 
and impose periodic boundary conditions, giving a matrix size of
$5 N^2$.
 In our model, we take $t=0.2$ eV, $\varepsilon_{d}-\varepsilon_{\pi^*} = 1.5$ 
eV, $U=1$ eV. 
The value of $J_{AF} S^2$ varies between $0.01-0.1$ eV. 
For each Monte-Carlo step,  we calculate
eigenvalues and eigenvectors of the Hamiltonian
with direct diagonalization. The Coulomb repulsion on the TCNE molecule is 
calculated within the Hartree-Fock limit,
\begin{equation}
 \langle U \rangle = \sum_j \left<\Psi_0|U n_{j\uparrow} 
n_{j\downarrow}|\Psi_0 \right>,
\end{equation}
where $|\Psi_0 \rangle$ is the lowest many-body state of $H_0$
obtained by filling the eigenstates up to the chemical potential using
the Fermi-Dirac function for a finite temperature.
In our Monte-Carlo 
 calculations, for a given number of valence electrons, we 
calculate the 
average magnetization $\left<|{\bf M}|\right>$ of the whole lattice and 
average angles of core spins, $\langle \theta_i \rangle$,  at
different values of $J_{AF}$.  For average magnetization  calculations, we only
consider  the contribution of the manganese core spins.
In our
calculations, we take 8000-32000 Monte-Carlo steps in total, sufficiently
larger than the 1000-5000  Monte-Carlo steps 
needed to establish the equilibrium state. We choose
a low temperature of 5 K to minimize the temperature fluctuations.

\begin{figure}[t]
\begin{center}
\includegraphics[angle=0,totalheight=1.9in]{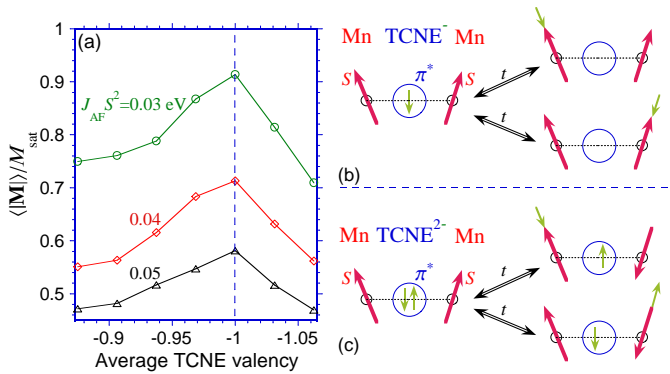}
\caption{\label{magnetization} 
(color online) 
(a) The normalized magnetization $\langle |{\bf M}|\rangle/M_{\rm sat}$, where 
$M_{\rm sat}$ is the saturation magnetization, as a function of the 
average TCNE valency for  superexchange coupling strengths
$J_{AF}S^2=0.03, 0.04$, and 0.05 eV. (b) Schematic showing
 how the delocalization of an unpaired electron stabilizes
the  ferromagnetic exchange. The left side shows
the lowest configuration; the right side shows the hopping processes that
stabilize the magnetic coupling.
(c) Schematic showing
how the delocalization of two $\pi^*$ electrons stabilizes
the  antiferromagnetic exchange.}
\end{center}
\end{figure}

 The central result is shown in Fig. \ref{magnetization}(a), 
where we see a significant change in the magnetization 
when electrons are transferred from manganese to TCNE
for different values of the superexchange coupling. If 
not all the TCNE are ionized in the initial state, illuminination 
with light creates a charge-transfer excitation from the Mn core spin
to the valence band consisting of the TCNE $\pi^*$ states and the 
Mn $3d$ states antiparallel to the core spin. The magnetization increases 
up to a formal valency of $-1$ for TCNE. The number of electrons in the
valence band can be further increased by exciting with light having a smaller 
energy creating divalent TCNE molecules. 
However, these excitations lead to a decrease 
in magnetization, see Fig. \ref{magnetization}(a).

\begin{figure}[t]
\begin{center}
\includegraphics[angle=0,scale=0.6]{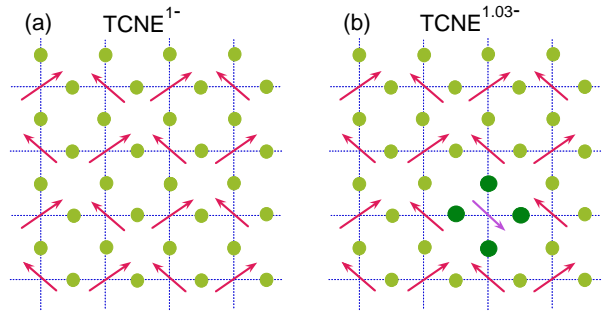}
\caption{\label{spins}
(color online) 
(a) Canted spin state for an average TCNE valency of $-1$ 
(the actual electron density on TCNE is 0.97). The 
arrows indicate the manganese core spins. The circles indicate 
the TCNE positions. (b) Spin state for an average TCNE valency of $-1.03$. 
Note that
the excited electron is delocalized over the TCNE molecules surrounding
the flipped spin, where the average density is 1.2 electrons (indicated 
by the larger dark-green circles).
 }
\end{center}
\end{figure}

To understand the trends in magnetization, we have to distinguish
the regions where the average TCNE valency is less and greater than $-1$.
Figure \ref{magnetization}(b) shows schematically how two manganese 
core spins are coupled ferromagnetically by an unpaired $\pi^*$ electron.
For parallel core spins, the conduction electron can lower its kinetic
energy by coupling to the empty Mn $t_{2g}$ states. 
To obtain a ferromagnetic ground state, it is important to include the 
Coulomb interaction on the TCNE ions. This can be understood as follows.
In the ferromagnetic state,
the coupling to the manganese core spins splits the TCNE states in
a spin-up and spin-down band. For an average TCNE valency 
of $-1$, the spin-down 
band (antiparallel to the core spins) is full, and 
there is no kinetic energy gain. 
However, for an antiferromagnetic configuration of manganese spins,
kinetic energy can still be gained since the TCNE $\pi^*$ orbital
can be doubly occupied. Therefore, this state is lower in energy 
when the  average electron density per TCNE is greater than $\frac{1}{2}$, 
even in the
absence of a superexchange interaction between the core spins.
However, the antiferromagnetic  state becomes unfavorable 
in the presence of a Coulomb interaction on the TCNE. 

\begin{figure}[t]
\begin{center}
\includegraphics[angle=0,totalheight=2.0in]{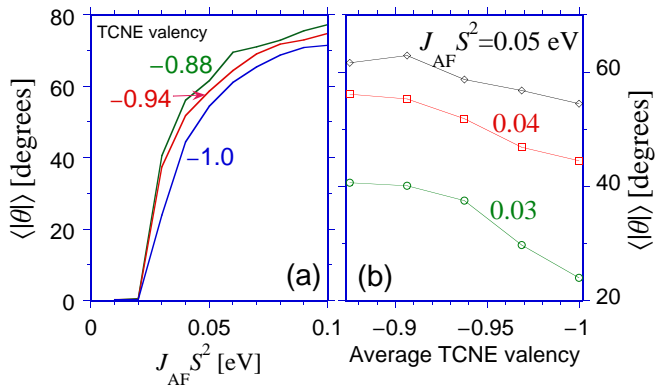}
\caption{\label{canting} 
(a) The average canting $\langle |\theta|\rangle$ 
with respect to the magnetization axis as a function of the 
strength of the superexchange coupling $J_{\rm AF}S^2$. (b)
The average canting $\langle |\theta|\rangle$ as a function of
the average TCNE valency for $J_{\rm AF}S^2=$0.03, 0.04, and 0.05 eV. }
\end{center}
\end{figure}

The behavior as a function of TCNE valency is a result of the competition
between the ferromagnetic coupling  mediated
by the unpaired TCNE $\pi^*$ electrons and the antiferromagnetic 
superexchange through the occupied TCNE $\pi$ orbitals.
Our Monte-Carlo simulations show that a canted spin state 
is formed.
A typical example is shown in  Fig. \ref{spins}(a) for $J_{\rm AF}S^2$=0.05 eV. 
To study the average canting angle of the Mn core spins
with respect to the magnetization
axis, we define 
$\langle |\theta| \rangle = \frac{1}{N^2}\sum_i \langle |\theta_i |\rangle$. 
Figure \ref{canting}(a) shows  $\langle |\theta| \rangle$ as a function
of the strength of the superexchange $J_{\rm AF}S^2$. We see that canting 
occurs  for a critical superexchange coupling, here $J_{AF}S^2= 0.02$ eV,
see Fig. \ref{canting}(a). Of interest for the photoinduced effect is the
change in canting angle as a function of the average TCNE valency.
From Fig. \ref{canting}(b), we see that a change in TCNE valency of 
0.1 electrons can cause a change in the average canting angle of about 
10-20$^\circ$. An increase in the number of unpaired electrons on TCNE
leads
to a stronger ferromagnetic  coupling and hence a decrease of the canting 
angle. Therefore, photoinduced transitions for electron densities
less than one electron per TCNE, increase the magnetization, 
see Fig. \ref{magnetization}(a).

The situation drastically changes when charge-transfer excitations are made with a 
smaller energy, where optical transitions are made from the manganese core spin
to the TCNE, creating a doubly occupied $\pi^*$ orbital. This changes the 
coupling between the manganese spins from ferromagnetic to antiferromagnetic,
as is shown schematically in Fig. \ref{magnetization}(c). Instead 
of letting one $\pi^*$ electron hybridize with both neighboring manganese 
atoms, it is kinetically more advantageous to flip one of the manganese
spins allowing both $\pi^*$ electrons to hybridize. For a larger system, 
the excited electron delocalizes over the TCNE molecules surrounding
the flipped spin, see Fig. \ref{spins}(b). Note that the average valency
of the TCNE surrounding the flipped spin is not $-2$, but closer to 
$-1.2$, explaining why divalent TCNE molecules are not observed experimentally
in, e.g., M\"ossbauer spectroscopy \cite{mossbauer}.

Summarizing, we have described a mechanism for photoinduced magnetism
in Mn-TCNE systems. Illumination with visible light causes 
charge-transfer transitions
from the Mn core spins to the valence shell consisting of 
the Mn $3d$ states antiparallel to the core spins and the TCNE $\pi^*$ 
orbitals. When the  average density of electrons per TCNE
is less than one, photoexciting 
electrons into the valence states strengthens the double-exchange 
coupling between the Mn core spins. This decreases the canting 
angle and increases the magnetization. Illumination with a larger 
wavelength creates divalent TCNE molecules. The magnetic coupling 
dramatically changes and locally core spins are flipped, causing a decrease
in magnetization. The $S'=2$ core spin  created by the 
photoexcitation is sensitive to  Jahn-Teller distortions and forms a local
polaron. Polaron formation can strongly
influence the dynamics of the system and can provide a mechanism to
explain the long lifetime of the photoinduced magnetic state.

The authors acknowledge Ken Ahn, A. J. Epstein, and J. S. Miller for 
fruitful discussions. This work was  supported by 
 the U.S. Department of Energy (DE-FG02-03ER46097), Research Cooperation
and NIU's Institute
for Nanoscience, Engineering, and Technology under a grant from the U.S.
Department of Education. Work at
Argonne National Laboratory was supported by the U.S. Department of 
Energy, Office of Basic Energy Sciences, under contract 
W-31-109-ENG-38.




\begin{thebibliography}{10}

\bibitem{zutic} I. Zutic, J. Fabian, and S. Das Sarma, 
Rev. Mod. Phys. {\bf 76}, 323 (2004). 

\bibitem{exp1} Y. Ogawa, S. Koshihara, K. Koshino, T. Ogawa, C. Urano,
 and H. Takagi,  Phys. Rev. Lett. {\bf 84}, 3181 (2000).

\bibitem{exp2} S. Koshihara {\it et al.}, 
Phys. Rev. Lett. {\bf 78}, 4617 (1997). 

\bibitem{exp3} K. Matsuda, A. Machida, Y. Moritomo, and A. Nakamura,  
Phys. Rev. B {\bf 58}, 4203 (1997).

\bibitem{cofe1} O. Sato, T. Iyoda, A. Fujishima, and K. Hashimoto, 
Science {\bf 272}, 704 (1996).

\bibitem{cofe2}  D. A. Pejakovic, J. L. Manson, J. S. Miller, and 
A. J. Epstein, Phys. Rev. Lett. {\bf 85}, 1994 (2000). 

\bibitem{mntcne}  D. A.
Pejakovic, C. Kitamura, J. S. Miller and A. J. Epstein, Phys. Rev. Lett. 
{\bf 88}, 057202 (2002).

\bibitem{theorycofe} T. Kawamoto, Y, Asai, and S. Abe, Phys. Rev. Lett. 
{\bf 86}, 348 (2001).

\bibitem{mossbauer} J. Zhang, J. Ensling, V. Ksenofontov, P. Gutlich, 
A. J. Epstein and J. S. Miller,  
Angew. Chem. Int. Ed. Engl. {\bf 37}, 657 (1998).

\bibitem{reentrant} C. M. Wynn, M. A. Girtu, J. Zhang, J. S. Miller, 
and A. J. Epstein, Phys. Rev. B {\bf 58}, 8508 (1998).

\bibitem{struct} J. S. Miller and A. J. Epstein, Chem. Commun. {\bf 1998}, 1319.  

\bibitem{gaussian} 
Gaussian 03, Revision C.02, M. J. Frisch, {\it et al.} 
Gaussian, Inc., Wallingford CT, 2004.
                                                                                
\bibitem{lsda} J. C. Slater, Quantum Theory of Molecular and Solids. 
Vol. 
4: The Self-Consistent Field for Molecular and Solids (McGraw-Hill, New 
York, 1974); S. H. Vosko, L. Wilk, and M. Nusair, Can. J. Phys. 58, 1200 
(1980).

\bibitem{pcm} V. Barone and M. Cossi, J. Phys. Chem. A {\bf 102}, 1995, 
(1998).

\bibitem{pimpolaron}  D. A.
Pejakovic, C. Kitamura, J. S. Miller, and A. J. Epstein, J. Appl. Phys.
{\bf 91}, 7176 (2002).

\bibitem{de1} C. Zener, Phys. Rev. {\bf 82}, 403 (1951).

\bibitem{de2}  P. W. Anderson and H. Hasegawa, Phys. Rev. {\bf 100}, 
675 (1955). 

\bibitem{de3} P.-G. de Gennes, Phys. Rev. {\bf 118}, 141 (1960). 

\end{thebibliography}
\end{document}